\documentclass[a4paper,twoside,10pt,twocolumn]{article}%
\usepackage[compress]{cite}
\usepackage{geometry}
\usepackage{authblk}
\usepackage{multirow}
\usepackage{url}
\usepackage{cuted}

\usepackage{hyperref}
\usepackage{breakurl}

\usepackage{placeins}
\usepackage{amsmath}%
\usepackage{amsfonts}%
\usepackage{amssymb}%
\usepackage{graphicx}
\usepackage{german}
\usepackage[ngerman,english]{babel}
\usepackage[T1]{fontenc}
\usepackage[utf8]{inputenc}
\usepackage{aeguill}

\begin{document}
\title{\textit{In-situ} Polarized Neutron Reflectometry: Epitaxial Thin Film Growth of Fe on Cu(001) by DC Magnetron Sputtering}
\author[1]{Wolfgang Kreuzpaintner}
\author[1]{Birgit Wiedemann}
\author[3]{Jochen Stahn}
\author[4]{Jean-Fran\c{c}ois Moulin}
\author[1]{Sina Mayr}
\author[2]{Thomas Mairoser}
\author[2]{Andreas Schmehl}
\author[2]{Alexander Herrnberger}
\author[3]{Panagiotis Korelis}
\author[4]{Martin Haese}
\author[1]{Jingfan Ye}
\author[4]{Matthias Pomm}
\author[1]{Peter B\"oni}
\author[5]{Jochen Mannhart}

\affil[1]{Technische Universit\"at M\"unchen, Physik-Department E21, James-Franck-Str. 1, 85748 Garching, Germany}
\affil[2]{Zentrum f\"ur elektronische Korrelation und Magnetismus, Universit\"at Augsburg, Lehrstuhl f\"ur Experimentalphysik VI, Universit\"atsstr. 1, 86159 Augsburg, Germany}
\affil[3]{Laboratory for Neutron Scattering and Imaging, Paul Scherrer Institut, 5232 Villigen PSI, Switzerland}
\affil[4]{Helmholtz-Zentrum Geesthacht, Zentrum f\"ur Material und K\"ustenforschung, Au\ss enstelle am MLZ in Garching bei M\"unchen, Lichtenbergstr. 1, 85748 Garching, Germany}
\affil[5]{Max-Planck-Institut f\"ur Festk\"orperforschung, Heisenbergstr. 1, 70569 Stuttgart, Germany}

\date{\today}
\maketitle
\begin{strip}
\begin{abstract}
The step-wise growth of epitaxial Fe on Cu(001)/Si(001), investigated by \textit{in-situ} polarized neutron reflectometry is presented. A sputter deposition system was integrated into the neutron reflectometer AMOR at the Swiss neutron spallation source SINQ, which enables the analysis of the microstructure and magnetic moments during all deposition steps of the Fe layer. We report on the progressive evolution of the accessible parameters describing the microstructure and the magnetic properties of the Fe film, which reproduce known features and extend our knowledge on the behavior of ultrathin iron films.
\end{abstract}
\end{strip}

\section{Introduction}

Owing to their valuable electronic, magnetic, and optical properties, thin films and heterostructures are indispensable in scientific and technological applications and offer fascinating prospects for the realization of advanced electronic devices \cite{1,2,4,5,6,7,8,12,13,14,15}. As a result, an increasing number of thin films and heterostructures are grown with atomic-layer precision by means of physical vapour deposition from complex materials \cite{52}. The material spectrum also broadens; sophisticated heterostructures of high complexity use a steadily increasing number of elements of the periodic table \cite{54,55}. At the same time, the control of defects and intended sample properties becomes more relevant. As morphologies, including sample structure, stoichiometry and defect population evolve with the deposition, so do the magnetic properties of the sample. It is, hence, highly desirable -- and even more challenging -- to analyze both as a function of layer thickness \textit{in-situ}. While the \textit{in-situ} characterization of films by electron- and photon-based probes \cite{56,57} as well as by scanning probe techniques \cite{58,59} is common practice, only a few attempts have been made to characterize the emerging sample properties by neutron scattering \cite{17,18,53}. 
However, as a spin-sensitive technique, Polarized Neutron Reflectometry (PNR) is very sensitive to both, structural and magnetic properties with atomic resolution. It is, hence, well established as an indispensable \textit{ex-situ} method to investigate samples in their final state. If it were possible to routinely perform PNR \textit{in-situ} on growing films and heterostructures, PNR would be even more valuable, as it can contribute to answering the grand questions of how the microstructure, defects, and if applicable, magnetic properties of heterostructures i) form, ii) are correlated with each other, and iii) how they evolve during growth. The results will be particularly valuable because all PNR data is accumulated from the very same sample.

As neutron sources and neutron optical concepts have strongly evolved in the last decades, and with data storage densities approaching regimes where a fundamental understanding of magnetism on the atomic scale is the key for further progress, today \textit{in-situ} PNR (\textit{i}PNR) appears as forthcoming analytical technique. 

Therefore, we decided to investigate the current state of viability and the potential of \textit{i}PNR in the context of analyzing the progressive evolution of the accessible microstructure parameters and the magnetic properties of a sputter deposited epitaxial Fe film on a Cu(001)$_\text{45\,\text{nm}}$/Si(001) {substrate}. This sample type was specifically chosen as its structural and accompanying magnetic properties have been widely studied in the past for different deposition and analysis techniques on a variety of substrates \cite{19,20,21,22,23,24,25}, yet, only little work has been done on sputter deposited Fe thin films \cite{26}. Any potentially different growth mode and a deviating magnetic behavior of sputtered films could provide both, more insight into the physics of thin magnetic films and a benchmarking of the \textit{i}PNR method.

As today’s neutron sources do not yet provide the required brilliance for \textit{in-operando} PNR experiments, the data presented in the following were taken while the coating process was periodically interrupted for the \textit{i}PNR measurements. In order to avoid potential surface contamination, special attention was given to a compatibility of the vacuum quality of our \textit{in-situ} thin film deposition setup with the required neutron data acquisition times. The coating setup offers a base pressure of $5.0 \times 10^{-9}$\,mbar, which was created by a turbo molecular pump (TMP). Due to the TMP's working principle of momentum transfer, the main constituent of the residual gas in the vacuum chamber is H$_2$, which only weakly interacts with the Fe surface. Contaminating residual gas species are typically two orders of magnitude below H$_2$, such that a monolayer formation time of $\sim 10^4$ - $10^5$\,s can be assumed. To further rule out any contaminating influences from residual gas species, we aimed at reducing our \textit{i}PNR data acquisition times to the lowest possible value by combining our \textit{in-situ} deposition setup with the prototype of the focusing Selene neutron optical concept \cite{37,37a}. It uses a pair of Montel-mirrors to focus a broad-wavelength-band neutron beam onto the sample and is capable of providing the data within 15\,min per spin direction for our \textit{i}PNR measurements. The data acquisition times were therefore sufficiently fast to avoid any relevant contamination of the Fe surface before the next Fe deposition step was performed.

\section{Experimental Procedure}
\subsection{\textit{In-situ} Thin Film Preparation}
\begin{figure}[ht]
	\centering
		\includegraphics[width=\columnwidth]{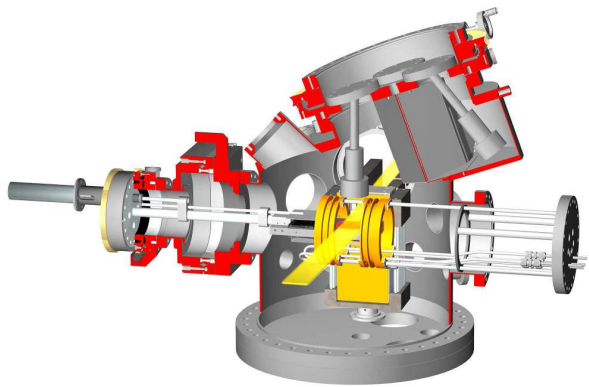}
			\caption{Schematic cross section of the \textit{in-situ} sputter deposition chamber: the sample manipulator is mounted on the left flange. The sample, exposed to the neutron beam (sketched in yellow) is located in the centre. A retractable pair of Helmholtz-coils used to magnetize the sample is mounted at the right flange. For magnetic fields exceeding 30\,mT, the coils are replaced by permanent magnets. The three sputter sources are located on the top, separated by shields to avoid cross contamination.}
			\label{fig:Kreuzpaintner_001}
\end{figure}

The coating setup is equipped with three 2\,inch sputter deposition sources, which were operated in direct current mode. The sputter guns are implemented such that either of the sputter sources can be rotated to a position perpendicular to the sample surface.
A schematic cross section and details of the sputtering system are shown in figure \ref{fig:Kreuzpaintner_001}. 
In-vacuum guide fields were implemented to maintain the neutron polarization up to the sample position. Stepping motors on linear and rotary vacuum feed-throughs are used to align the sample in the neutron beam. A more detailed design description of the deposition setup will be presented elsewhere.
The thin films were deposited epitaxially \textit{in-situ} in the neutron beam using metal-metal-epitaxy-on-silicon (MMES) \cite{27,28,29,30,32,34}.
After a 45\,nm thick Cu(001) seed layer, a 7.0\,nm thick Fe layer was grown in 28 separate deposition steps $i$ from a 99.99\% pure Fe sputter target at an Ar sputtering gas pressure of $4.50 \times 10^{-3}$\,mbar. The DC sputtering power of 20\,W resulted in a deposition rate of $0.18$\,$\mu$g\,cm$^{-2}$\,s$^{-1}$. The deposition of the equivalent of approximately 1 monolayer of Fe per deposition step was controlled by the opening times of a deposition shutter (typically 1.5\,s per deposition step). Between two deposition steps the chamber was evacuated to base pressure and the \textit{i}PNR measurements were carried out. After the 14$^\text{th}$ Fe deposition step, the \textit{i}PNR measurement were only performed after every second coating step.

\subsection{\textit{In-situ} Polarized Neutron Reflectometry}
The unique feature of the AMOR beamline at PSI is that most components are mounted on an optical bench. The instrument is, therefore, highly flexible and allows both, the installation of the \textit{in-situ} sputter deposition chamber and the insertion of the prototype of the Selene neutron guide \cite{37,37a}. It ends 400\,mm before the focal point and is fully compatible with the deposition setup, where the distance from the fused silica (SiO$_2$) neutron window of the \textit{in-situ} deposition setup to the sample is 380\,mm. 

\begin{figure}[ht]
	\centering
		\includegraphics[width=0.8\columnwidth]{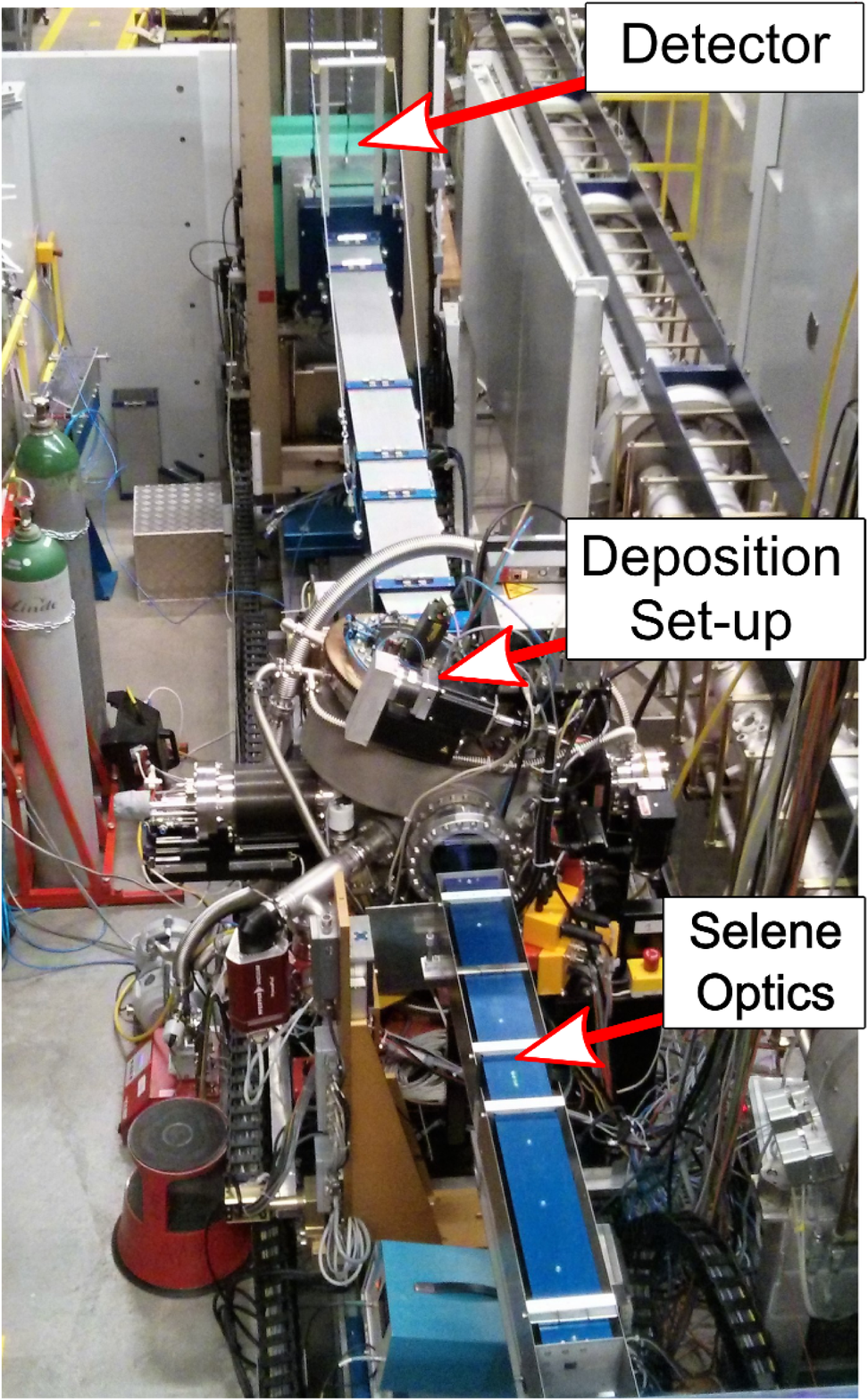}
			\caption{Sputter deposition chamber and Selene optics, integrated into the AMOR beamline: The \textit{in-situ} chamber is located in the centre of the beam. Clearly visible is the neutron window made from fused silica (SiO$_\textit{2}$). The neutrons leave the vacuum chamber on the opposite side through an identical window, followed by a flight tube made from aluminum and the two dimensional detector (turquoise box). The blue Selene guide elements are located in front of the sputtering chamber.}
			\label{fig:Kreuzpaintner_002}
\end{figure}

Figure \ref{fig:Kreuzpaintner_002} shows the integration of the coating set-up and the Selene optics into AMOR. The sputter process was controlled remotely.
For the \textit{i}PNR measurements a magnetic field of 70\,mT was applied to the sample perpendicular to the scattering plane using permanent magnets. Since in the Selene mode the complete beam is convergent and the sample is in the focal point, no further beam-shaping elements between the optics and the sample were needed and the full beam divergence of 1.6\,$^{\circ}$ was used to illuminate the sample with a neutron wavelength band of 4 -- 10\,\AA. This leads to a gain factor of 30 when compared to the conventional PNR operation mode of AMOR. However, the resolution in $\frac{\Delta q_z}{q_z}$ becomes $q_z$-dependent (see \cite{63} for details). With the settings applied for our measurements, the resolution quickly increases from $\frac{\Delta q}{q} \approx 4.5\%$ in the regime of total reflection to a quasi-stable value of $\frac{\Delta q}{q} \approx 2.3\%$ for $q_z \gtrsim 0.2$\,nm$^{-1}$. Beam polarization was realized by the transmittance of the neutrons through an $m = 4.2$ Fe/Si multilayer polarizer with a logarithmic spiral shape. The neutron polarization was selected by an RF spin-flipper.

\section{Results and Discussion}
\begin{figure}[h!]
	\centering
		\includegraphics[width=1\columnwidth]{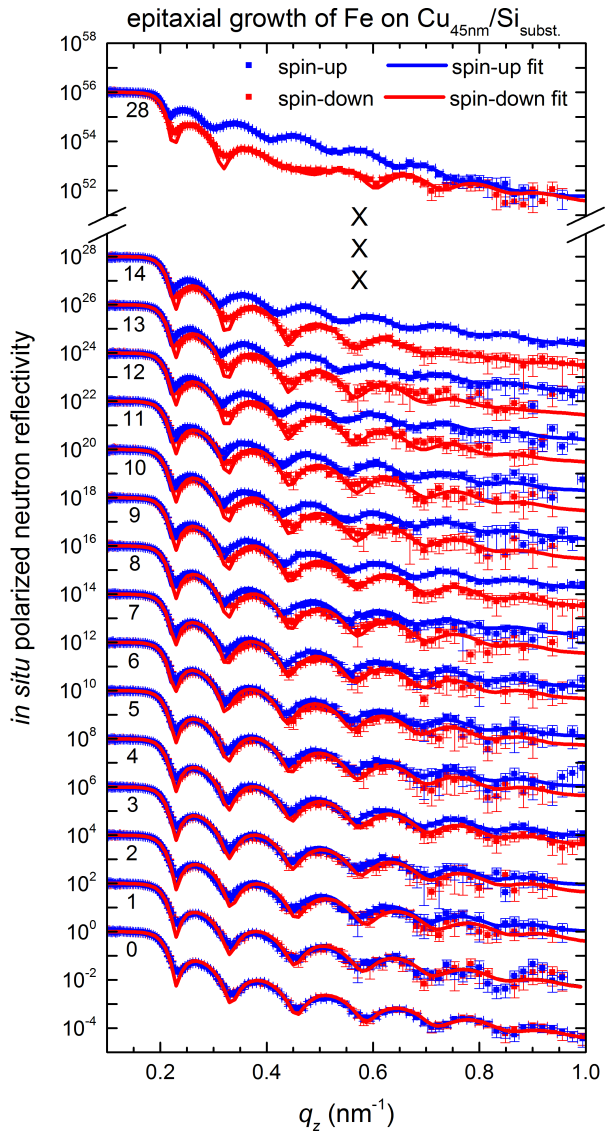}
			\caption{Measured \textit{i}PNR data, overlaid with fitted reflectivity curves: Shown is the neutron reflectivity versus the perpendicular momentum transfer $q_z$. Each pair of curves is vertically shifted by two orders of magnitude for better visibility. Typically the \textit{i}PNR data acquisition time was 15 minutes per spin direction. The number below the regime of total reflection denotes the deposition step $i$ of the epitaxial Fe layer. Each deposition step between the \textit{i}PNR measurements required approximately 5 minutes.}
			\label{fig:Kreuzpaintner_003}
\end{figure}

The \textit{i}PNR data overlaid with the fitted reflectivity curves is shown in figure \ref{fig:Kreuzpaintner_003}. Each pair is characterized by four key parameters: a) the critical edge up to which total reflection occurs, revealing the scattering length density from which the number density of each layer ($n^{\text{Cu}}$ and $n^{\text{Fe}}$)  is obtained; b) the periodic Kiessig fringes, a measure of the layer thickness $d^{\text{Cu}}$ and $d^{\text{Fe}}$; c) the decay of the reflectivity curves that exceeds the expected decrease in the Fresnel reflectivity, a measure for the interfacial root-mean-square (rms) roughness $\sigma^{\text{Cu/Si}}$ on the Cu/Si and $\sigma^{\text{Fe/Cu}}$ on the Fe/Cu interfaces; and d) the splitting of the spin-up (+) and spin-down (-) reflectivity $R^+$ and $R^-$, providing quantitative information on the magnetic moments in the sample.

Whilst for deposition step $i=1$ $R^+$ and $R^-$ are identical, the gradual increase in the splitting between $R^+$ and $R^-$ from $i=2$ to $i=28$ directly correlates with the magnitude of the in-plane magnetization $M^{\text{Fe}}$ of the Fe layer and with $d^{\text{Fe}}$. The \textit{i}PNR data was analyzed quantitatively using the SimulReflec Software Package \cite{38} assuming a two layer model: Fe on Cu$_\text{seed}$ on Si substrate. The parameters of the Cu layer, i.\,e.\ $d^\text{Cu}=45.14 \left( +0.21 \atop -0.14 \right)$\,nm, $n^\text{Cu}=8.36 \left( +0.19 \atop -0.11 \right) \times 10^{22}$\,cm$^{-3}$ and $\sigma^\text{Cu/Fe} = 0.63\left( +0.12 \atop -0.18 \right)$\,nm were kept constant while the parameters of the Fe layer were varied. 

\begin{figure*}[h!]
	\centering
		\includegraphics[width=0.7\textwidth]{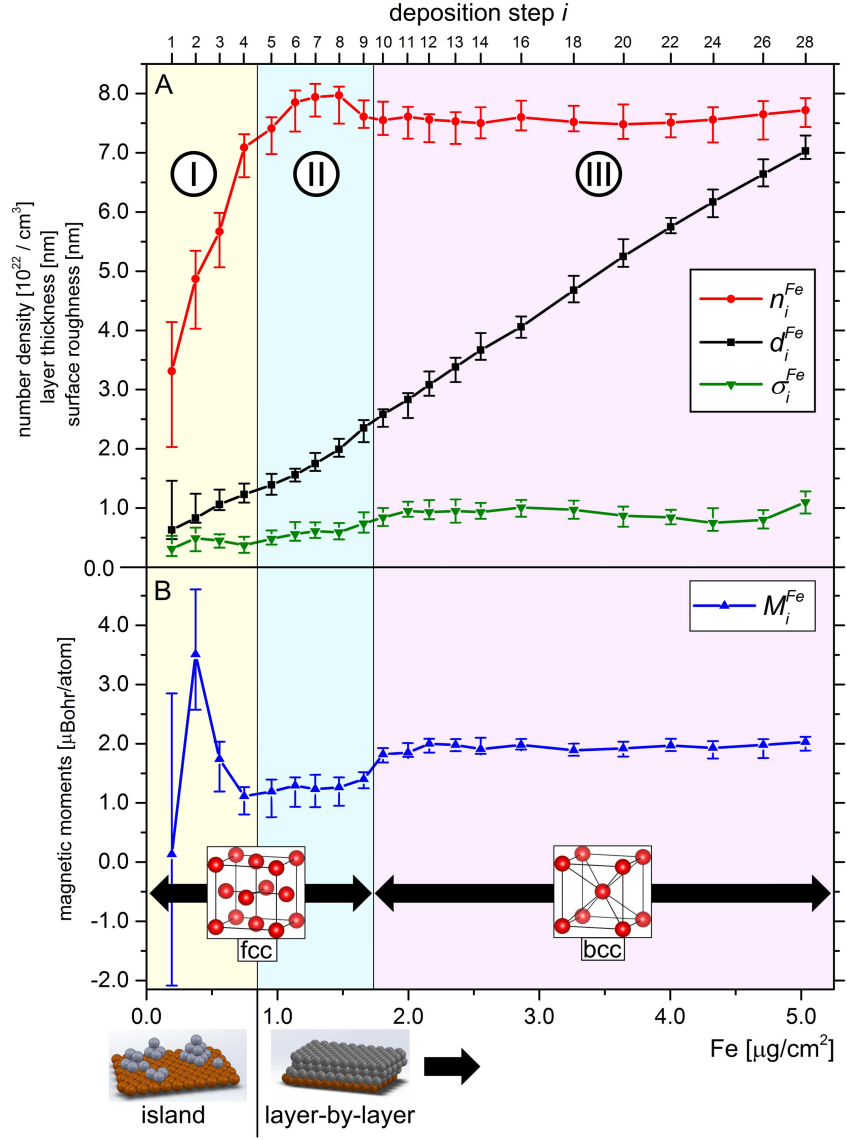}
			\caption{The fit parameters of the epitaxially grown Fe layer. Three main regimes (I--III) with different characteristic behaviors for the number density $n^{\text{Fe}}_i$, thickness $d^{\text{Fe}}_i$, interfacial roughness $\sigma^{\text{Fe}}_i$ and magnetization $M^{\text{Fe}}_i$ can be identified. Shown are also the concluded growth modes (island/layer-by-layer) and crystalline structures (fcc/bcc). 
			}
			\label{fig:Kreuzpaintner_004}
\end{figure*}

The resulting fit parameters $d^{\text{Fe}}_i$\,[nm], $n^{\text{Fe}}_i$\,[$10^{22}$\,cm$^{-3}$], $\sigma^{\text{Fe}}_i$\,[nm] and $M^{\text{Fe}}_i$\,[$\mu_\text{Bohr}$/atom] and their evolution are shown in figure \ref{fig:Kreuzpaintner_004} as a function of $i$ and the amount of deposited material. The errors of the Cu and Fe parameters are estimated by a 5\% increase over the optimum figure of merit $FOM \sim \sum{\left|\ln{R_{\text{fit}}}-\ln{R_{\text{meas}}}\right|}$ on independent variation of a single parameter \cite{68}, 
where $R_{\text{fit}}$ is the fitted and $R_{\text{meas}}$ the measured reflectivity, respectively. 

Three regimes I -- III in the evolution of the fit parameters can be identified:
\paragraph{Regime I:\\}
Deposition step $i=1$ generates an Fe layer with an apparent thickness of $d^{\text{Fe}}_1=0.63$\,nm (approximately three monolayers) and a very low number density of Fe atoms: $n^{\text{Fe}}_1=3.31\times 10^{22}$/cm$^3$ if compared to the bulk value ($\sim 8.48 \times 10^{22}$/cm$^3$). This density can only be rationalized by requiring the scattering length density of the layer to be composed of a weighted average of the scattering lengths of Fe ($b^{\text{Fe}}=3.31$\,fm) and vacuum ($b^{\text{vac}}=0$\,fm). The low density implies that either the first three monolayers form islands or a layer of very small density. 

The data of the following deposition steps $2\leq i < 5$ indicates, too, that the film indeed started its growth in the island mode \cite{48,49}, because these steps yield only a small relative increase in thickness but a density $n^{\text{Fe}}_i$, which increases significantly faster than $d^{\text{Fe}}_i$. The simultaneously occurring reduction of surface roughness for steps $3\leq i\leq 5$ relative to the thickness of the layers also traces the coalescence of separate Fe islands.
Interestingly, the coating applied in deposition step $i=1$ shows an in-plane magnetization of a mere 0.13\,$\mu_{\text{Bohr}}$/atom, which we attribute to a strong perpendicular magnetic anisotropy \cite{41,42,43,44,49} or superparamagnetism of nanoscale islands \cite{60,61}.

While the density and the thickness of the layers increase continuously with each deposition step, the in-plane magnetization varies strongly. After $i=2$ ($d^{\text{Fe}}_2=0.83$\,nm) the film exhibits an in-plane magnetization of 3.5\,$\mu_\text{Bohr}$/atom. Ultrathin Fe layers on various substrates with magnetization exceeding the bulk level have been reported before \cite{43,45,64,65} and are confirmed by our measurements, yet the magnetization in our film in its nucleation phase might exceed even these large literature values of up to $\sim 3.1$\,$\mu_\text{Bohr}$/atom \cite{64,65}.
At deposition step $i=4$, the magnetization has decreased from its maximum ($i=2$) to a level of $\sim 1.25$\,$\mu_\text{Bohr}$/atom, where it remains approximately constant up to growth step $i=9$.

\paragraph{Regime II:\\}
After deposition step $i = 4$, the Fe islands have completely coalesced, as revealed by the change in increase in thickness from 0.95\,nm/($\mu$g\,cm$^{-2}$) for $3\leq i\leq 4$ to 1.35\,nm/($\mu$g\,cm$^{-2}$) for $i >4$, which coincides with the phasing-out of the increase in density ($4\leq i\leq 5$). A transition to a layer-by-layer growth with the 5$^\text{th}$ deposition step must, therefore, be concluded. The density of the Fe layer reaches a value of $\sim 7.95 \times 10^{22}$/cm$^3$ ($6\leq i\leq 8$).

The evolvement of the magnetism is directly visible in \textit{i}PNR by the clear separation of the spin-polarized raw data. Room-temperature magnetism of Fe thin films has previously only been reported for thicknesses below $\sim$ 4 \cite{41,42,43,44} and above 12 atomic layers \cite{44}. According to the literature, Fe films with a thickness of 5–11 atomic layers have a Curie temperature $T_C$ of only 275–280\,K \cite{44}. The increase in $T_C$ to above room temperature in our experiment falls in line with the enhanced magnetization of the film shown during its nucleation. We attribute the enhancement of the magnetization to the microstructure of the \textit{in-situ} grown films differing from the ones of the Fe films reported in the literature \cite{44} caused by the use of sputtering as deposition method.

An oscillatory magnetic behavior \cite{41,42,46}, resulting from antiferromagnetic coupling between single atomic Fe layers with intrinsic perpendicular magnetization could not be confirmed in our \textit{i}PNR measurements. It is noted that the applied magnetic field of 70\,mT may have been sufficiently strong to overcome the anisotropy, rotating the magnetic moments in-plane and thereby suppressing these oscillations \cite{49,62}.

\paragraph{Regime III:\\}
As the film continues to grow through steps $8\leq i\leq 12$, its number density decreases to $\sim 7.6\times 10^{22}$/cm$^3$. In parallel, the interfacial roughness increases slightly, and the magnetization increases from $\sim 1.25$\,$\mu_\text{Bohr}$/atom to $\sim 2$\,$\mu_\text{Bohr}$/atom. Along with deposition step $i=9$, the growth rate changes from 1.35\,nm/($\mu$g\,cm$^{-2}$) to 1.40\,nm/($\mu$g\,cm$^{-2}$). In their combination, these changes strongly indicate a magnetically driven phase transition from the face-centered-cubic (fcc) to the body-centered-cubic (bcc) phase that the Fe film undergoes at around $i = 9$. This phase transition is known to exist for Fe films with a thickness of $\sim 10 - 12$ atomic layers \cite{44,47,48}.

Growing further, all properties of the film stay remarkably constant. Its magnetization equals $\sim 2$\,$\mu_\text{Bohr}$/atom which is close to the bulk value of $\sim 2.2$\,$\mu_\text{Bohr}$/atom of Fe.

\section{Summary and Conclusions}
We have probed the magnetic and structural properties of a thin film of Fe that was epitaxially grown in UHV on a Cu(001)$_\text{45\,\text{nm}}$/Si(001) substrate using \textit{i}PNR. 
The combination of Montel optics with DC magnetron sputtering in UHV allowed the \textit{in-situ}  \ collection of spin-polarized neutron data during the sequence of 22 growth steps while keeping the sample fixed in the neutron beam. Avoiding any movements of the sample is ideal for detecting small variations in the \textit{i}PNR signal. Moreover, the analysis of the data is facilitated because it is based on one and the same sample.

Our \textit{i}PNR measurements confirm most of the known thickness dependent magnetic properties of Fe layers. However, we have observed some new features in our sputter deposited Fe layers when compared with layers grown by other techniques. These include an indication for a large magnetization during the early nucleation phase that exceeds the literature values \cite{64,65} by more than 10\%. We also observed magnetism at room temperature in films with a thickness of 5 to 11 atomic layers, which corresponds to an increase of $T_C$ of at least $20$\,K if compared to the $T_C$ of the Fe films reported in the literature \cite{44}.

While the understanding of the evolution of Fe films during their growth is of interest in itself, our studies simultaneously demonstrate the viability and potential of \textit{i}PNR for the analysis of magnetic properties on an atomic scale. Here, \textit{i}PNR can clearly provide relevant data that complements the data obtained from photon and electron-based techniques. In fact, the future prospects of \textit{i}PNR are tantalizing: possible scientific questions for \textit{i}PNR include the investigation of perpendicular magnetic anisotropy \cite{2}, magneto-elastic coupling \cite{51,4,5}, and magnetism at oxide interfaces and the corresponding topology \cite{6,7,8}. We expect that \textit{i}PNR will also be of great benefit in the investigation of the build-up of chirality or incommensurability \cite{9,10} or the formation of solitons/skyrmions in films or at interfaces \cite{14,15} during growth.

In addition, the detailed observation of the processes taking place during topotactic transformations \cite{1} or the formation of self organized structures \cite{11} can be followed up \textit{in-situ}. In this context, beamlines might, however, be preferable that in addition to reflectometry also allow large-angle scattering geometries, such that additionally to the data obtained by \textit{i}PNR crystal structures including defects like oxygen vacancies can be analyzed.

While we demonstrated the \textit{in-situ} technique using sputtering as deposition method, \textit{i}PNR will be equally well applicable for MBE or PLD, in particular because the deposition of adatoms on chamber walls can more easily be minimized, thereby avoiding neutron activation. Compared to the setup presented in this work, an implementation of \textit{i}PNR at the future European Spallation Source ESS  using the next generation Selene-optics (to be realized for the reflectometer ESTIA \cite{67} at ESS), the flux at the sample will increase by approximately a factor of 4000 thus decreasing the measuring time to below half a second for the two spin channels \cite{66}. Therefore, \textit{i}PNR will even provide sufficient time resolution for probing the structural and magnetic properties during thin film growth, both, \textit{in-situ} and \textit{in-operando} thus opening new fascinating applications in the field of thin films.


\section*{Acknowledgements}
This work is based on experiments performed on AMOR at the Swiss spallation source SINQ, Paul Scherrer Institut, Switzerland and upon experiments performed at the REFSANS instrument operated by HZG at the Heinz Maier-Leibnitz Zentrum (MLZ), Garching, Germany. Part of the work was supported by the Swiss National Science Foundation through the National Centre of Competence in Reasearch MaNEP and by the Deutsche Forschungsgemeinschaft via the Transregional Research Center TRR 80.

\small
\bibliography{Fe}{}

\begin{thebibliography}{10}
\bibitem{66}
The gain of 4000 at ESS when compared with SINQ is obtained as follows: i) 100
  times larger peak brilliance, 10 times larger beam area (20\,mm $\times$
  20\,mm), and a 4 times larger solid angle. Moreover, a larger range of
  momentum transfers $q_z$ is probed.
\end{thebibliography}
\bibliographystyle{unsrt}
\end{document}